\documentclass[12pt,a4paper]{article}

\pdfoutput=1

\usepackage[DIV13,BCOR0mm]{typearea}

\usepackage{natbib}
\bibpunct[, ]{(}{)}{,}{a}{}{,}

\usepackage{graphicx}

\newcommand{\bigfrac}[2]{\mbox{$\displaystyle\frac{#1}{#2}$}}
\begin{document}

\title{\textbf{Continuum-mechanical, Anisotropic Flow
               model for polar ice masses,
               based on an anisotropic
               Flow Enhancement factor}}

\author{\textsc{Luca Placidi}\thanks{E-mail: luca.placidi@uniroma1.it}\\[0.4ex]
        {\normalsize Department of Structural and Geotechnical Engineering, ``Sapienza'',}\\[-0.35ex]
        {\normalsize University of Rome, Via Eudossiana 18, I-00184 Rome, Italy}\\[0.4ex]
        {\normalsize Smart Materials and Structures Laboratory,}\\[-0.35ex]
        {\normalsize c/o Fondazione ``Tullio Levi-Civita'',}\\[-0.35ex]
        {\normalsize Palazzo Caetani (Ala Nord), I-04012 Cisterna di Latina, Italy}\\[1.4ex]
        \textsc{Ralf Greve}\\[-0.1ex]
        \textsc{Hakime Seddik}\\[0.4ex]
        {\normalsize Institute of Low Temperature Science, Hokkaido University,}\\[-0.35ex]
        {\normalsize Kita-19, Nishi-8, Kita-ku, Sapporo 060-0819, Japan}\\[1.4ex]
        \textsc{S\'ergio H. Faria}\\[0.4ex]
        {\normalsize GZG, Department of Crystallography, University of G\"ottingen,}\\[-0.35ex]
        {\normalsize Goldschmidtstra\ss{}e 1, D-37077 G\"ottingen, Germany}}

\date{}

\maketitle

\begin{abstract}

\vspace*{-171mm}

\noindent\hspace*{-10mm}\emph{Cont.~Mech.~Thermodyn.}
\textbf{22}~(3), 221--237 (2010).
doi: 10.1007/s00161-009-0126-0
\\{}\hspace*{-10mm}Authors' version;
original publication available at www.springerlink.com

\vspace*{162mm}

A complete theoretical presentation of the
Continuum-mechanical, Anisotropic Flow model, based on an
anisotropic Flow Enhancement factor (CAFFE model) is given. The
CAFFE model is an application of the theory of mixtures with
continuous diversity for the case of large polar ice masses in
which induced anisotropy occurs. The anisotropic response of
the polycrystalline ice is described by a generalization of
Glen's flow law, based on a scalar anisotropic enhancement
factor. The enhancement factor depends on the orientation mass
density, which is closely related to the orientation
distribution function and describes the distribution of grain
orientations (fabric). Fabric evolution is governed by the
orientation mass balance, which depends on four distinct
effects, interpreted as local rigid body rotation, grain
rotation, rotation recrystallization (polygonization) and grain
boundary migration (migration recrystallization), respectively.
It is proven that the flow law of the CAFFE model is truly
anisotropic despite the collinearity between the stress
deviator and stretching tensors.

\end{abstract}

\section{Introduction}
\label{sect_intro}

In order to study the mechanical behaviour of large polar ice
masses, we use the method of continuum mechanics. Generally,
ice is treated as an incompressible, isotropic and extremely
viscous non-Newtonian fluid, and Glen's flow law \citep{glen55,
nye_57} is used as a constitutive equation. However, for thick
polar ice masses anisotropic behaviour occurs, and therefore,
Glen's flow law must be changed. Many efforts have been
undertaken to deal with this problem \citep[e.g.][]{hutter83,
budd_jacka_89, jackabudd89, azuma95, staroszczykmorland01,
Morland2003, placidihutterfaria06_gamm, gagliardini_etal_09}.
In this paper we will present a continuum-mechanical model,
which is based on earlier work by
\citet{faria01,faria_london_i,faria_london_iii,placidi04_PhD,
placidi05_PhD, placidifariahutter04_annals,
placidihutter06_zamp, placidihutter06_cmt}. The model is
referred to as the Continuum-mechanical, Anisotropic Flow
model, based on an anisotropic Flow Enhancement factor, or
``CAFFE model'' for short.

The macroscopic anisotropy of polar ice is due to its
microstructure. Ice is a polycrystalline material made of a
vast number of crystallites (grains), the mechanical behaviour
of which is extremely anisotropic \citep{jackabudd89}. A single
ice crystal shows transversely isotropic behaviour, and its
$c$-axis (optical axis) defines the privileged direction
\citep{boehler87}. Slide along basal planes, orthogonal to the
$c$-axis, is easier than slide along other crystallographic
planes, and since the study by \citet{mcconnel91} it has been
common to refer to this as the deck-of-cards behaviour of ice.
However, the transition from the mechanics of a single crystal
to that of a huge polycrystal entails the complication of
different deformation mechanisms, and the selection of these
mechanisms in different situations. A continuum approach is
deemed appropriate in order to deal with this problem in a
manageable fashion. We choose the approach of describing the
polycrystalline ice as a mixture \citep{truesdell57a,
truesdell57b}, the species of which are the grains
characterized by a certain orientation \citep{faria_london_ii}.
The orientations of the crystallites, i.e., the unit vectors
parallel to the $c$-axes, belong to a continuous space, so that
the ice is considered a Mixture with Continuous Diversity (MCD)
\citep{faria01}.

In the MCD theory, equations are defined at the species level
(``microscopic'') and at the mixture level (``macroscopic'').
However, the ``microscopic'' equations do \emph{not} govern the
evolution of single crystallites, the dynamics of which is
hidden in the theory. The objective of the mixture approach is
to predict the polycrystalline behaviour only. In other words,
the CAFFE model is a macroscopic model, and the microstructure
is taken into account only phenomenologically, without going
down to the actual microscopic level. The same holds for
classical continuum mechanics: we know that matter is
discontinuous, spaces between particles (atoms, molecules,
etc.) are empty and the molecular structures have strong
influences on the mechanical behaviour. However, the
microstructural characteristics are not resolved in detail;
instead, constitutive equations for a continuous body are
postulated that are in accordance with experimental data and
general principles like determinism, objectivity and the Second
Law of Thermodynamics. In the same way, in a MCD, the behaviour
of a single species is important for the mixture dynamics, but
single species dynamics does not correspond to any measurable
quantity.

The general set of equations for polycrystalline ice modelled
as a MCD was developed by \citet{faria01, faria_london_i,
faria_london_ii, faria_london_iii, placidihutter06_cmt}, and
restrictions of the constitutive equations due to the Second
Law of Thermodynamics were given by these authors.
\citet{placidi04_PhD, placidi05_PhD, placidihutter06_zamp}
suggested explicit forms for the constitutive relations. In
this study, we present the CAFFE model as an improved version
of these previous formulations. After defining the notation
(Section~\ref{sect_notation}), we derive in
Section~\ref{sect_caffe} in a rational way the set of CAFFE
equations, while taking care that the following requirements
are satisfied:
\begin{itemize}
\item All fundamental principles of classical continuum
    mechanics must be fulfilled.
\item The model must be sufficiently simple to allow
    numerical implementation in current flow models for
    polar ice masses.
\item The parameters of the model must in principle be
    measurable by either laboratory experiments or field
    observations.
\end{itemize}
In Section~\ref{ssect_omd}, we define the orientation mass
density, and in Section~\ref{ssect_omb} we deal with the
general mass balance that governs its evolution. In
Section~\ref{ssect_omb_conrel}, we characterize the
constitutive quantities introduced in Section~\ref{ssect_omb}
in order to describe grain rotation, local rigid body rotation,
grain boundary migration (or migration recrystallization) and
polygonization (or rotation recrystallization). In
Sections~\ref{ssect_aniso_flow_law} and
\ref{ssect_aniso_flow_law_inv}, we present an anisotropic
generalization of Glen's flow law based on a scalar anisotropic
enhancement factor. It is similar to that by
\citet{placidihutter06_zamp}, but simpler and consistent with
the Second Law of Thermodynamics.

Section~\ref{sect_aniso_glen} is devoted to the analysis of the
anisotropic properties of the CAFFE flow law. We note the
collinearity between the stretching and the stress deviator
tensors and show that collinearity and isotropy do not share
any fundamental concepts, in the sense that non-collinear,
isotropic flow laws as well as collinear, anisotropic flow laws
are possible \citep[e.g.][]{liu02, faria08}. A discussion of
the advantages and disadvantages of collinearity in an
anisotropic flow law terminates Section~\ref{sect_aniso_glen}.

Some examples for the constitutive relations introduced in
Section~\ref{ssect_omb_conrel} are provided in the appendix
(Section~\ref{sect_omd_evol}).

\section{Notation}
\label{sect_notation}

For a general field $A$, the star superscript $A^{\ast}$
denotes an orientational dependence
$A^{\ast}\left(\mathbf{x},t,\mathbf{n}\right)$, where $t$ is
the time, $\mathbf{x}$ the position vector and $\mathbf{n}$ the
orientation (unit vector parallel to the $c$-axis) in the
present configuration. Otherwise the field
$A\left(\mathbf{x},t\right)$ is supposed to be independent of
$\mathbf{n}$. It is implicitly assumed that for a given
position $\mathbf{x}$ all possible orientations $\mathbf{n}$
are defined. The gradient ($\nabla$) and divergence
($\nabla\cdot$) operators are applied, as usual, to the space
variable $\mathbf{x}$, while the orientational gradient
($\nabla^\ast$) and divergence ($\nabla^\ast\cdot$) operators
are applied to the orientational variable $\mathbf{n}$. For
arbitrary scalars $A^{\ast}$ and vectors $\mathbf{A}^{\ast}$,
we define
\begin{eqnarray}
  \nabla A^{\ast} =\bigfrac{\partial A^{\ast}}{\partial \mathbf{x} }\,,
  & &
  \nabla^\ast A^{\ast}=\bigfrac{\partial A^{\ast}}{\partial \mathbf{n} }
  - \left( \bigfrac{\partial A^{\ast}}{\partial \mathbf{n} }
  \cdot \mathbf{n} \right) \mathbf{n}\,,
  \label{gradient_differential_operator}
  \\[1ex]
  \nabla \cdot \mathbf{A}^{\ast} = \mathrm{tr}\left[ \nabla
  \mathbf{A}^{\ast}\right] =\mathrm{tr}\left[ \bigfrac{\partial
  \mathbf{A}^{\ast}}{\partial \mathbf{x} }\right]\,,
  & &
  \nabla^\ast
  \cdot \mathbf{A}^{\ast}=\mathrm{tr} \left[  \bigfrac{\partial
  \mathbf{A}^{\ast}}{\partial \mathbf{n} } - \left(
  \bigfrac{\partial \mathbf{A}^{\ast}}{\partial \mathbf{n} } \cdot
  \mathbf{n} \right) \, \mathbf{n}\right]\,.
  \label{divergence_differential_operator}
\end{eqnarray}
An immediate consequence of
Eq.~(\ref{gradient_differential_operator})$_2$ is $\nabla^\ast
A^{\ast}\cdot\mathbf{n}=0$. We also note that the explicit form
of the orientational gradient operator in spherical coordinates
is
\begin{eqnarray}
  \nabla^\ast A^{\ast} &=& \mathbf{e}_1 \left[ \cos \theta \cos
  \varphi \bigfrac{\partial A^{\ast} }{\partial \theta }
  -\bigfrac{\sin \varphi }{\sin \theta }
  \bigfrac{\partial A^{\ast} }{\partial \varphi }\right]
  \nonumber\\
  &+& \mathbf{e}_2 \left[ \cos \theta \sin \varphi \bigfrac{\partial
  A^{\ast} }{\partial \theta }+\bigfrac{\cos \varphi }{\sin \theta
  }\bigfrac{\partial A^{\ast} }{\partial \varphi }\right] +
  \mathbf{e}_3 \left[ -\sin \theta \bigfrac{\partial A^{\ast}
  }{\partial \theta }\right]\,,
  \label{3du5}
\end{eqnarray}
where
$\left\lbrace\mathbf{e}_1,\mathbf{e}_2,\mathbf{e}_3\right\rbrace$
is a fixed orthonormal basis on which we project vectors and
tensors, and $\theta$ and $\varphi$ are the zenith and azimuth
angle, respectively. The orientation $\mathbf{n}$ can be
parameterized as follows,
\begin{equation}
  \mathbf{n}=\left(
  \begin{array}{c}
  \sin \theta \cos \varphi  \\
  \sin \theta \sin \varphi  \\
  \cos \theta
  \end{array}
  \right)
  = \sin \theta \cos \varphi \,\, \mathbf{e}_1
    + \sin \theta \sin \varphi \,\, \mathbf{e}_2
    + \cos \theta \,\, \mathbf{e}_3\,.
  \label{sphericalvectorial}
\end{equation}

\section{CAFFE model}
\label{sect_caffe}

\subsection{Orientation mass density, orientation distribution function}
\label{ssect_omd}

\begin{figure}[htb]
  \centering
  \includegraphics[angle=0,width=0.5\textwidth]{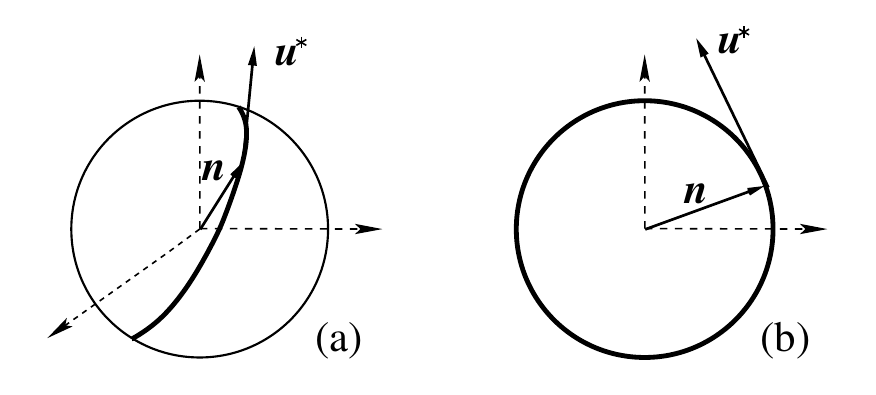}
  \caption{(a) Each point of the unit sphere $\mathcal{S}^2$
  represents a particular orientation $\mathbf{n}$. The orientation
  transition rate $\mathbf{u}^{\ast}$ is orthogonal
  to $\mathbf{n}$. Panel (b) shows a projection on the plane spanned by
  $\mathbf{u}^{\ast}$ and $\mathbf{n}$. The anticlockwise
  direction of the rotation corresponds to a spin velocity
  $\mathbf{s}^{\ast}$ pointing out of the plane.}
  \label{ssquareandu}
\end{figure}

In the CAFFE model, each point of the continuous body is
interpreted as a representative volume element of the
polycrystal that encloses a large number of crystallites with
their own orientations. Each orientation is represented by a
unit vector $\mathbf{n}\in\mathcal{S}^{2}$ (where
$\mathcal{S}^{2}$ denotes the unit sphere) parallel to the
$c$-axis (Fig.~\ref{ssquareandu}a).

In the MCD framework, distributions of continuous species
parameters (like the orientation) are expressed in terms of
associated mass densities. This means that for every time $t$
and position $\mathbf{x}$ an \emph{orientation mass density}
$\varrho^{\ast}\left(\mathbf{x},t,\mathbf{n}\right)$ is defined
such that, when integrated over $\mathcal{S}^{2}$, the usual
mass density of the polycrystal
$\varrho\left(\mathbf{x},t\right)$ results,
\begin{equation}
  \varrho \left( \mathbf{x},t\right)
  = \int\limits_{\mathcal{S}^{2}}\!\varrho ^{\ast
  }\left( \mathbf{x},t,\mathbf{n}\right)
  \,\mathrm{d}^{2}n\,,
  \label{mass_density_definition}
\end{equation}
where $\mathrm{d}^{2}n$
($=\sin{\theta}\,\mathrm{d}\theta\,\mathrm{d}\varphi$ in
spherical coordinates) is the solid angle increment on the unit
sphere $\mathcal{S}^{2}$. The orientation mass density
$\varrho^{\ast}$, as stated in
Eq.~(\ref{mass_density_definition}), has the following physical
meaning: the product
$\varrho^{\ast}\!\left(\mathbf{x},t,\mathbf{n}\right)\,\mathrm{d}^{2}n$
is the mass fraction of crystallites with orientations directed
towards $\mathbf{n}$ within the solid angle $\mathrm{d}^{2}n$.
Therefore, the quantity $\varrho^{\ast}/\varrho$ can be
interpreted as the analogue of the usual orientation
distribution function (ODF) \citep[e.g.][]{rashid92,
placidihutter04_warsaw}, which is also used in the context of
liquid crystals \citep[e.g.][]{Blenk1992, Papenfuss2000} or in
mesoscopic damage mechanics \citep[e.g.][]{Massart2004719,
Papenfuss2008}. However, we remark that in the glaciological
community, the ODF usually describes the relative number, and
not the mass fraction, of grains with a certain orientation.

\subsection{Orientation mass balance}
\label{ssect_omb}

Some kinematic quantities are required in order to describe the
evolution of positions and orientations. We will use the
classical velocity $\mathbf{v}\left(\mathbf{x},t\right)$ and
the orientation transition rate
$\mathbf{u}^{\ast}\left(\mathbf{x},t,\mathbf{n}\right)$ as
kinematic rates. The velocity $\mathbf{v}$ represents the
transition of mass from a given position to a neighbouring
position in three-dimensional space. Analogously, the
orientation transition rate $\mathbf{u}^{\ast}$ represents the
transition of mass from a certain orientation to a neighbouring
orientation on the unit sphere. Note that the velocity is
assumed to be independent on the orientation, while the
orientation transition rate can depend on the orientation [for
a longer discussion on this topic see e.g.\
\citet{faria_london_i}].

As shown by \citet{faria01}, the normality condition
$\mathbf{n}\cdot\mathbf{n}=1$ makes the orientation transition
rate $\mathbf{u}^{\ast}$ orthogonal to $\mathbf{n}$
($\mathbf{u}^{\ast}\cdot\mathbf{n}=0$) and to the spin velocity
$\mathbf{s}^{\ast}$ of the crystallites
($\mathbf{u}^{\ast}=\mathbf{s}^{\ast}\times\mathbf{n}$); see
also the caption of Fig.~\ref{ssquareandu}b. In the MCD theory,
the balance of mass is formulated as
\begin{equation}
  \bigfrac{\partial \varrho^{\ast}}{\partial t} +\nabla \cdot \left[
  \varrho^{\ast}\mathbf{v} \right] +\nabla^\ast \cdot \left[
  \varrho^{\ast}\mathbf{u}^{\ast} \right ]
  = \varrho^{\ast} \Gamma^{\ast}\,.
  \label{micro_mass_balance}
\end{equation}
The quantity $\Gamma^{\ast}$ is the specific mass production
rate. It describes the rate of change of mass (per unit mass)
of one species (for crystallites having a certain orientation
$\mathbf{n}$) into another species with a different
orientation. This corresponds physically to the effect of grain
boundary migration (migration recrystallization).

Integration of Eq.~(\ref{micro_mass_balance}) over the unit
sphere $\mathcal{S}^2$ gives the balance of mass of the
polycrystal,
\begin{equation}
  \bigfrac{\partial \varrho}{\partial t}
  + \nabla \cdot [\varrho\mathbf{v}] = 0\,,
  \label{macro_mass_balance}
\end{equation}
provided that Eq.~(\ref{mass_density_definition}) and
\begin{equation}
  \int_{\mathcal{S}^2} \varrho^{\ast} \Gamma^{\ast} \,\mathrm{d}^2 n = 0,
  \qquad
  \int_{\mathcal{S}^2} \nabla^\ast \cdot [\varrho^{\ast}\mathbf{u}^{\ast}]
  \,\mathrm{d}^2 n = 0
  \label{micro_macro_1}
\end{equation}
hold. Note that Eq.~(\ref{micro_macro_1})$_1$ describes the
conservation of mass, and Eq.~(\ref{micro_macro_1})$_2$ is a
consequence of the Gauss theorem.

\subsection{Constitutive equations for the orientation mass balance}
\label{ssect_omb_conrel}

\begin{figure}[htb]
  \centering
  \includegraphics[angle=0,width=0.65\textwidth]{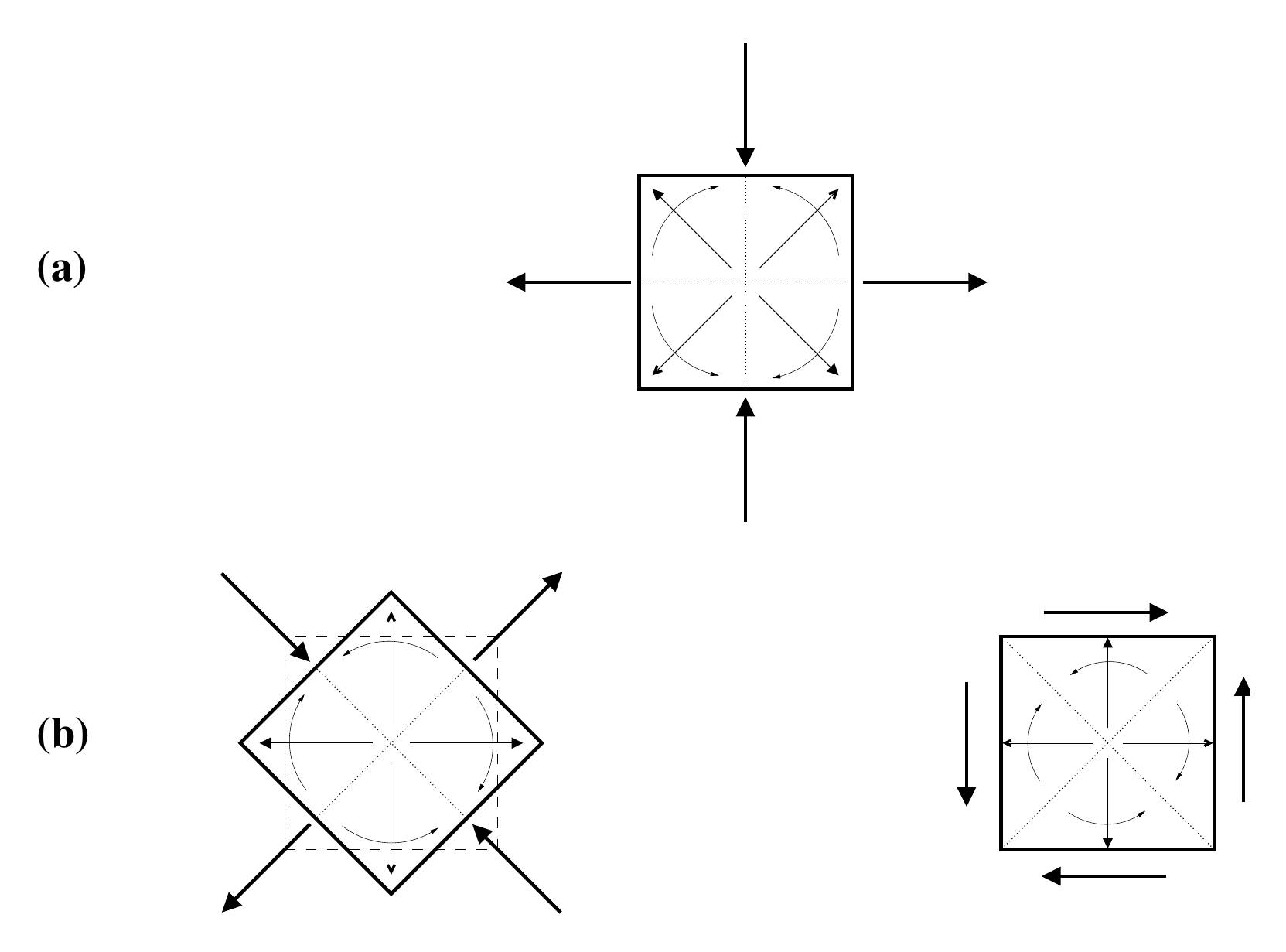}
  \caption{Schematic, two-dimensional representations of representative
  volume elements. Straight arrows inside the volume elements show
  $c$-axes of crystallites well oriented for deformation. Curved
  arrows indicate the rotational motion of such crystallites due to
  grain rotation (``deck-of-cards effect''). Arrows outside the volume
  elements mark the state of stretching.
  In case (a), the state of stretching is pure shear.
  The $c$-axes are well oriented at angles of $45^{\circ}$ with
  respect to the axes of extension and compression, and the $c$-axes
  rotate towards the axis of compression and away from the axis of
  extension. Case (b) shows the same configuration rotated anticlockwise
  by $45^{\circ}$ (rotated pure shear) for two equivalent volume elements.}
  \label{rotrec}
\end{figure}

\citet{placidihutter06_cmt} showed that the orientation
transition rate $\mathbf{u}^{\ast}$ can be decomposed into two
parts,
\begin{equation}
  \mathbf{u}^{\ast}
  = \mathbf{u}^{\ast}_\mathrm{rbr} + \mathbf{u}^{\ast}_\mathrm{c}
  = \mathbf{W}\cdot\mathbf{n} + \mathbf{u}^{\ast}_\mathrm{c}\,,
  \label{decomp_u}
\end{equation}
where
$\mathbf{u}^{\ast}_\mathrm{rbr}=\mathbf{W}\cdot\mathbf{n}$ is
the contribution of the local rigid body rotation of the
polycrystal, $\mathbf{W}=\mathrm{Skw}\,\mathbf{L}$ is the
skew-symmetric part of the velocity gradient
$\mathbf{L}=\nabla\mathbf{v}$ and
$\mathbf{u}^{\ast}_\mathrm{c}$ is a vector to be constitutively
prescribed. For the latter, we introduce phenomenologically a
further decomposition,
\begin{equation}
  \mathbf{u}_\mathrm{c}^{\ast}
  = \mathbf{u}^{\ast}_\mathrm{gr}
    + \frac{\mathbf{q}^{\ast}}{\varrho^{\ast}}\,,
  \label{decomp_uc}
\end{equation}
in order to separate the physical effects of (i) grain rotation
(``deck-of-cards effect'', see also Fig.~\ref{rotrec}), to be
modelled by $\mathbf{u}^{\ast}_\mathrm{gr}$, and (ii) rotation
recrystallization (polygonization), to be modelled by the
orientation flux $\mathbf{q}^{\ast}$ \citep{goedert_03}.

By assuming a linear dependence on the stretching tensor
$\mathbf{D}=\mathrm{Sym}\,\mathbf{L}$, the constitutive vector
$\mathbf{u}^{\ast}_\mathrm{gr}$ takes the form
\begin{equation}
  \mathbf{u}^{\ast}_\mathrm{gr}
  = \iota \, [ ( ( \mathbf{D} \cdot
  \mathbf{n} ) \cdot \mathbf{n} ) \mathbf{n}-\mathbf{D}
  \cdot \mathbf{n} ]\,,
  \label{constpart_u}
\end{equation}
where $\iota>0$ is a material parameter [$\iota$ is called
``shape factor'' in the theory of rotational diffusion; see
e.g.\ \citet{faria01}]. \citet{dafalias01} noted that, for the
special case $\iota=1$, the unit vector $\mathbf{n}$ remains
orthogonal to the associated material area element, and thus
the rotation is an \emph{affine rotation}. However, the
possibility of non-affine rotations cannot \emph{a priori} be
ruled out. An advantage of the present macroscopic approach is
the possibility to parameterize $\iota$ without any conflicts
with ``microscopic'' assumptions. [As a contrasting example,
the model by \citet{staroszczykmorland01} is also a macroscopic
model, but it is restricted to affine rotations.] In fact,
\citet{placidi04_PhD} showed that the fabrics in the upper 2000
m of the GRIP ice core in central Greenland can be best
explained by the value $\iota\approx{}0.4$. A study on the
EPICA ice core in Dronning Maud Land, East Antarctica, provided
a best fit between modelled and measured fabrics for
$\iota=0.6$ (Seddik et al. 2008).

The constitutive equations (\ref{decomp_u}) and
(\ref{constpart_u}) are not new in the literature and not
specific to ice \citep[e.g.][and references
therein]{larson1988, Blenk1992, Larson1999, Papenfuss2000,
dafalias01}. They were derived by \citet{placidi04_PhD} within
the MCD framework. We remark that, even though the unit vector
$\mathbf{n}$ specifying the orientation of the crystals is
unique, Eq.~(\ref{constpart_u}) is not the most general case.
In fact, \citet{dafalias01} discussed  the case of non-affine
rotations. More generally, \citet{faria01} and
\citet{placidihutter06_cmt} derived the thermodynamically
consistent class of these constitutive equations.

Following the argumentation by \citet{goedert_03}, the
orientation flux (which is supposed to describe rotation
recrystallization) is modelled as a diffusive process,
\begin{equation}
  \mathbf{q}^\ast
  = -\lambda\,\nabla\!^\ast
     [\rho^\ast \mathcal{H}^\ast]\,,
  \label{eq_ofl}
\end{equation}
where the parameter $\lambda>0$ is the orientation diffusivity,
and $\mathcal{H}^\ast$ is an orientation-dependent ``hardness''
function. However, recent results by \citet{durand_etal_08}
suggest that rotation recrystallization is an isotropic process
not affected by the orientation. In this case, the choice
\begin{equation}
  \mathcal{H}^\ast \equiv 1
  \label{eq_hardness_2}
\end{equation}
is indicated, which renders Eq.~(\ref{eq_ofl}) equivalent to
Fick's laws of diffusion on the unit sphere. We remark that in
the MCD theory the hardness function is called the chemical
potential for the given species. It is a constitutive quantity
that depends at least on the OMD $\varrho^{\ast}$.
Consequently, it is possible to show by applying the chain rule
that often an equivalent of Fick's law results even when
$\mathcal{H}^\ast$ is not a constant \citep[e.g.][]{faria01}.

\begin{figure}[htb]
  \centering
  \includegraphics[width=60mm,angle=0]{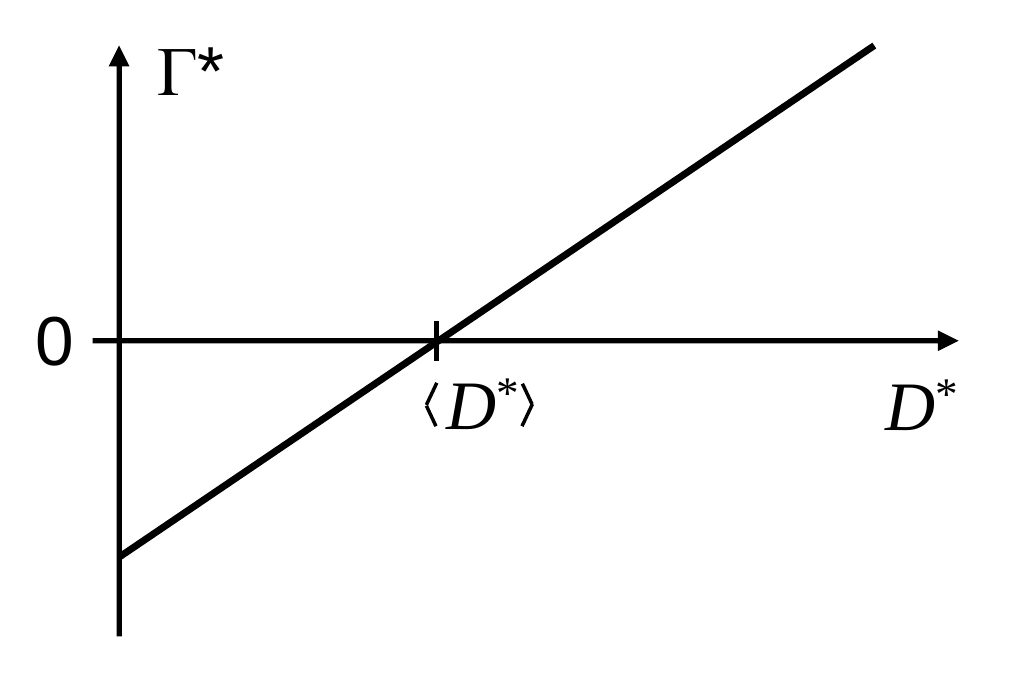}
  \par\vspace*{-2ex}\par
  \caption{Specific mass production rate $\Gamma^{\ast}$ according to
  Eq.~(\ref{Gammafunc2}).}
  \label{fig_opr}
\end{figure}

As for the specific mass production rate $\Gamma^{\ast}$, in
the studies by \citet{placidi04_PhD} and \citet{placidi05_PhD}
it was shown that, in order to model the effect of grain
boundary migration, a reasonable constitutive equation for
$\Gamma^{\ast}$ is
\begin{equation}
  \Gamma^{\ast}
  = \hat{\Gamma} \left[ D^{\ast}- \langle D^{\ast} \rangle \right]\,.
  \label{Gammafunc2}
\end{equation}
The dimensionless quantity $D^{\ast}$ is called the stretching
deformability of crystallites,
\begin{equation}
  D^{\ast} \equiv
  5\,\frac{(\mathbf{D}\cdot\mathbf{n})^2
  -\left( (\mathbf{D}\cdot\mathbf{n}) \cdot \mathbf{n} \right)^2}
  {\mathrm{tr}\left(\mathbf{D}^2\right)}\,;
  \label{Dcdefinition}
\end{equation}
its physical meaning is the square of the resolved shear strain
rate (or stretching) on the basal plane, normalized by the
orientation-independent scalar invariant
$\mathrm{tr}(\mathbf{D}^2)$. The additional factor $5$ is
merely a convention, for which the reason will become clear
below (Section~\ref{ssect_aniso_flow_law}). Further,
$\langle\cdot\rangle$ is the averaging operator
\begin{equation}
  \langle \cdot \rangle
  \equiv \int_ {\mathcal{S}^{2}} \frac{\varrho ^{\ast}}{\varrho }
  \,\left(\cdot\right) \,\mathrm{d}^2 n\,,
  \label{Gammafunc3}
\end{equation}
and $\hat{\Gamma}$ is a constitutive parameter (see
Fig.~\ref{fig_opr}). The conservation of mass expressed by
Eq.~(\ref{micro_macro_1})$_1$ is compatible with
Eqs.~(\ref{Gammafunc2})--(\ref{Gammafunc3}), and they are also
compatible with the rational constitutive theory developed by
\citet{placidihutter06_cmt}. Provided that $\hat{\Gamma}>0$,
Eqs.~(\ref{Gammafunc2})--(\ref{Gammafunc3}) have the effect
that $\Gamma^{\ast}$ is greater than zero when the stretching
deformability $D^{\ast}$ is high, and $\Gamma^{\ast}$ is less
than zero when the stretching deformability $D^{\ast}$ is low.
This means that crystallites well-oriented for deformation will
be enlarged \citep[e.g.][]{kamb72}. Since ice crystallites
deform essentially by basal shearing, the resolved shear rate
(which is proportional to the orientation dependence of
$D^{\ast}$) is related to the rate of accumulation of
deformation energy in the material, which drives dynamic
recrystallization.

The first contribution in Eq.~(\ref{decomp_u}),
$\mathbf{u}^{\ast}_\mathrm{rbr}$, is thermodynamically
reversible, because there is no energy dissipation associated
with local rigid body rotations. The second contribution,
$\mathbf{u}^{\ast}_\mathrm{c}$, has been split up in
Eq.~(\ref{decomp_uc}) into $\mathbf{u}^{\ast}_\mathrm{gr}$ and
$\mathbf{q}^\ast$. The grain-rotation part,
$\mathbf{u}^{\ast}_\mathrm{gr}$, is thermodynamically
irreversible because it is linearly dependent on the stretching
tensor $\mathbf{D}$, see Eq.~(\ref{constpart_u}), which by
definition must vanish for any reversible process in a viscous
fluid \citep[e.g.,][]{hutter83, mueller85, faria01}. The
rotation-recrystallization part, $\mathbf{q}^\ast$, is
thermodynamically irreversible because of the diffusive nature
of the process as specified in Eq.~(\ref{eq_ofl}). The specific
mass production rate $\Gamma^{\ast}$ is also thermodynamically
irreversible, because grain growth and recrystallization are
thermally activated, irreversible processes [see also the
discussion by \citet[][Sect.~3c]{faria_london_ii}]. Note that,
besides the physical interpretation, the thermodynamic
reversibility or irreversibility of these terms can also be
investigated by exploiting the Second Law of Thermodynamics
\citep{faria_london_i, placidihutter06_cmt}.

A problem is that it is not possible at the moment to constrain
the values of the two parameters $\hat{\Gamma}$ and $\lambda$
in a reasonable fashion. Determining these parameters by
experiments is very difficult, because the relevant time scales
are far too large and the strain rates far too low to be
reproduced in the laboratory. Deformation experiments, even if
conducted over a period of years, inevitably end up by
activating non-natural deformation and recrystallization
mechanisms. The only promising way out of this is to measure
the fabrics ($c$-axis distributions) and the changes in grain
stereology (sizes and shapes) in natural polar ice and fit
$\hat{\Gamma}$ and $\lambda$ to these observations. The
situation is complicated further by the fact that a functional
dependence of these parameters on temperature and/or
dislocation density should be considered
\citep[cf.][]{faria_london_i,faria_london_iii}. This requires
further attention.

Some examples for the orientation transition rate (grain
rotation) and the orientation production rate
(recrystallization) under different deformation regimes are
given in the appendix (Section~\ref{sect_omd_evol}).

\subsection{Anisotropic flow law}
\label{ssect_aniso_flow_law}

The anisotropic flow law presented by
\citet{placidihutter06_zamp} has been modified in order to make
it simpler and compatible with the Second Law of
Thermodynamics,
\begin{equation}
  \mathbf{D}
  = A \, \hat{E}(S)
    \left(\frac{\mathrm{tr}(\mathbf{S}^2)}{2}\right)^{(n-1)/2}
    \mathbf{S}\,,
  \label{new_flow_law}
\end{equation}
where $A$ and $n$ are the same rate factor and stress exponent,
respectively, as in the isotropic Glen flow law, $\mathbf{S}$
is the stress deviator defined by
\begin{equation}
  \mathbf{S}
  = \mathbf{t}
    -\left(\frac{\mathrm{tr}\,\mathbf{t}}{3}\right)\,\mathbf{I}
\end{equation}
(i.e., the deviatoric part of the Cauchy stress tensor
$\mathbf{t}$) and $\mathbf{I}$ is the identity tensor. Further,
$S\in{}[0,\,5/2]$ is the positive scalar
\begin{equation}
 S = \langle S^{\ast} \rangle
   =  \int_ {\mathcal{S}^{2}} \frac{\varrho^{\ast}}{\varrho} \,S^{\ast}
      \,\mathrm{d}^2 n\,,
  \label{stress_deformability}
\end{equation}
and $S^{\ast}$ is the analogue of $D^{\ast}$,
\begin{equation}
  S^{\ast} \equiv 5\,\frac{(\mathbf{S}\cdot\mathbf{n})^2
  -\left( (\mathbf{S}\cdot\mathbf{n})
  \cdot \mathbf{n} \right)^2 }
  {\mathrm{tr}(\mathbf{S}^2)}\,,
  \label{Dcdefinition2}
\end{equation}
which can also be written in terms of the Cauchy stress tensor
as
\begin{equation}
  S^{\ast}
  = 5 \,\frac{((\mathbf{t}\cdot\mathbf{n}) \times \mathbf{n})^2}
             {\mathrm{tr}(\mathbf{t}^2)}>0\,.
  \label{Dcdefinition2bis}
\end{equation}
We call $S^{\ast}$ the stress deformability of crystallites and
$S$ the stress deformability of the polycrystal. In the
literature, the scalar $(\mathbf{t}\cdot\mathbf{n})^2-
\left((\mathbf{t}\cdot\mathbf{n})\cdot\mathbf{n}\right)^2=
(\mathbf{S}\cdot\mathbf{n})^2-
\left((\mathbf{S}\cdot\mathbf{n})\cdot\mathbf{n}\right)^2$ has
been identified with the square of the resolved stress on the
basal plane, so that the stress deformability of crystallites,
Eq.~(\ref{Dcdefinition2}), can also be called the normalized
square of the resolved stress on the basal plane. As for the
stress exponent, it is often chosen as $n=3$
\citep[e.g.][]{paterson_94}, but we will keep it general in the
following.

For a thermodynamicist it may appear strange to formulate a
constitutive equation in terms of the stretching tensor and not
in terms of the stress deviator. However, there is no
inconsistency in this formulation. From the theoretical point
of view the stress is indeed the constitutive property and the
strain rate (stretching) is the variable, but whether we choose
this or the inverse relation is just a matter of taste or
custom. In the glaciological community the inverse form is most
commonly used; in the book by \citet{hutter83} the historical
reason for this is given.

Our new formulation of the flow law is not only compatible with
the Second Law of Thermodynamics \citep{placidihutter06_cmt},
but Eq.~(\ref{new_flow_law}) is much more flexible than the
previous Placidi--Hutter formulation. The mechanical
non-linearity
$\left(\mathrm{tr}(\mathbf{S}^2)/2\right)^{(n-1)/2}$ and the
anisotropic part $\hat{E}(S)$ are now nicely separated, so that
the choice of the stress exponent is not limited to $n=3$ any
more, and the new formulation is not even restricted to a power
law.

In Section~\ref{ssect_aniso_flow_law_inv} we will show that,
due to Eq.~(\ref{new_flow_law}), the two quantities $D^{\ast}$
and $S^{\ast}$ defined in Eqs.~(\ref{Dcdefinition}) and
(\ref{Dcdefinition2}) are identical,
\begin{equation}
  S^{\ast} = D^{\ast}\,,
  \label{deform_equal}
\end{equation}
so that we will simply call them the species (or
``crystallite'') deformability. In the same way, the positive
scalars $S$ and $D$ will be called the polycrystal
deformability,
\begin{equation}
  S = \langle S^{\ast} \rangle = \langle D^{\ast} \rangle = D\,.
  \label{poldef}
\end{equation}
Both the crystal and polycrystal deformabilities can only
assume values in the range from zero to 5/2,
\begin{equation}
  S^\ast = D^\ast \in [0,\,\mbox{$\frac{5}{2}$}]\,,
  \quad
  S      = D      \in [0,\,\mbox{$\frac{5}{2}$}]\,.
\end{equation}
The proof (which is laborious and shall not be detailed here)
involves to insert Eq.~(\ref{sphericalvectorial}) in
Eqs.~(\ref{Dcdefinition}) and (\ref{Gammafunc3}), and study the
maxima and minima of the deformabilities $D^\ast$ and $D$ for
general stretching tensors $\mathbf{D}$ as functions of the
zenith angle $\theta$ and the azimuth angle $\phi$. Taking into
account that for a randomly distributed OMD (isotropic fabric)
\begin{equation}
  \varrho^{\ast} = \frac{\varrho}{4\pi}
  \quad\Rightarrow\quad
  S = D = 1
  \label{S_iso}
\end{equation}
holds, the function $\hat{E}(S)$ is demanded to be monotone, of
class $C^1[0,\,5/2]$ and has the fixed points
\begin{equation}
  \hat{E}(0) = E_\mathrm{min}\,,\qquad
  \hat{E}(1) = 1\,,\qquad
  \hat{E}(\mbox{$\frac{5}{2}$})=E_\mathrm{max}\,,
  \label{enh_def}
\end{equation}
where $E_\mathrm{min}<1$ and $E_\mathrm{max}>1$ are the minimum
and the maximum enhancement factors. This means that if the
polycrystal deformability $S$ is highest ($S=5/2$), the flow
law (\ref{new_flow_law}) gives the maximum stretching, if the
polycrystal deformability $S$ is lowest ($S=0$), the flow law
(\ref{new_flow_law}) gives the minimum stretching, and if the
polycrystal deformability is the same as for the isotropic case
($S=1$), the flow law (\ref{new_flow_law}) reproduces the
classical Glen flow law.

\begin{figure}[htb]
  \centering
  \includegraphics[width=70mm]{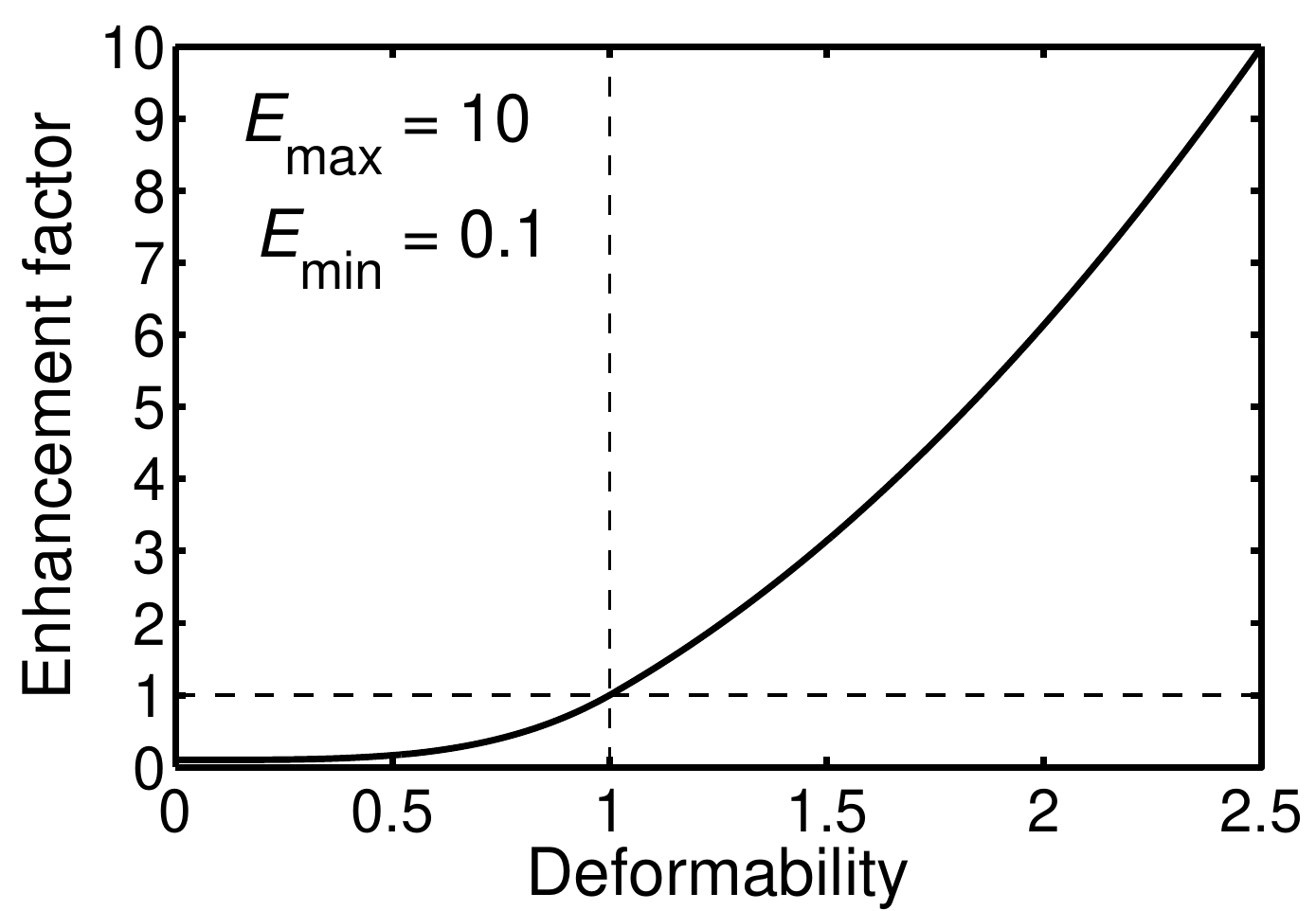}
  \caption{Anisotropic enhancement factor $\hat{E}(S)$ as
  a function of the deformability $S$ according to
  Eq.~(\ref{an_glen5}), for $E_\mathrm{max}=10$ and
  $E_\mathrm{min}=0.1$.}
  \label{fig_aniso_enh}
\end{figure}

As for the detailed functional form of the anisotropic
enhancement factor $\hat{E}(S)$, some experimental data suggest
that the enhancement factor depends on the ``averaged Schmid
factor'' to the fourth power \citep{azuma95, miyamoto99}. Since
the polycrystal deformability $S$ of the CAFFE model is related
to the square of the averaged Schmidt factor, it is reasonable
to assume a dependency of $\hat{E}$ on $S^2$. However, this
does not allow to fulfill Eq.~(\ref{enh_def}) for arbitrary
choices of the parameters $E_\mathrm{min}$ and
$E_\mathrm{max}$. Hence the function $\hat{E}(S)$ is chosen to
depend on $S^2$ in the interval $[1,\,5/2]$ [in which the
experiments by \citet{azuma95} and \citet{miyamoto99} have been
carried out] only, and for the interval $[0,1]$ a dependency on
$S^\tau$ is introduced. The exponent $\tau$ is adjusted such
that the function is continuously differentiable at $S=1$. This
yields
\begin{equation}
  \hat{E}\left( S \right)=\left\lbrace
  \begin{array}{lr}
  \left(1-E_\mathrm{min}\right)S^{\tau} + E_\mathrm{min}\,,\qquad
  \tau = \bigfrac{8}{21}\left(
  \bigfrac{E_\mathrm{max}-1}{1-E_\mathrm{min}} \right)\,,
  & S \in \left[0,1 \right]\,,   \vspace{1mm}   \\
  \bigfrac{4 S^2 \left( E_\mathrm{max}-1 \right)+25 - 4
  E_\mathrm{max}}{21}\,, & S \in \left[1, \frac{5}{2} \right]\,.
  \end{array}
  \right.
  \label{an_glen5}
\end{equation}
Several investigations \citep[e.g.][]{russell_head_budd_79,
pimienta_etal_87, budd_jacka_89} indicate that the parameter
$E_\mathrm{max}$ (maximum softening) is $\sim{}10$. The
parameter $E_\mathrm{min}$ (maximum hardening) can be
realistically chosen between 0 and $\sim{}0.1$, a non-zero
value serving mainly the purpose of avoiding numerical
problems. The function (\ref{an_glen5}) is shown in
Fig.~\ref{fig_aniso_enh}.

\subsection{Inversion of the anisotropic flow law}
\label{ssect_aniso_flow_law_inv}

The anisotropic flow law (\ref{new_flow_law}) can be inverted
analytically as long as the power-law form is retained. From
Eq.~(\ref{new_flow_law}) we find
\begin{equation}
  \frac{\mathrm{tr}(\mathbf{D}^2)}{2}
  = A^2\,\hat{E}^2(S)
    \left(\frac{\mathrm{tr}(\mathbf{S}^2)}{2}\right)^{n-1}
    \frac{\mathrm{tr}(\mathbf{S}^2)}{2}\,,
\end{equation}
and thus
\begin{equation}
  \frac{\mathrm{tr}(\mathbf{S}^2)}{2}
  = A^{-2/n}\,\hat{E}^{-2/n}(S)
    \left(\frac{\mathrm{tr}(\mathbf{D}^2)}{2}\right)^{1/n}\,.
\end{equation}
Inserting this result in Eq.~(\ref{new_flow_law}) and solving
for $\mathbf{S}$ yields
\begin{equation}
  \mathbf{S} = A^{-1/n}\,\hat{E}^{-1/n}(S)\,
               \left(\frac{\mathrm{tr}(\mathbf{D}^2)}{2}
               \right)^{-(1-1/n)/2}
               \mathbf{D}\,.
  \label{inv_new_flow_law_1}
\end{equation}

In order to complete the inversion, we must prove
Eqs.~(\ref{deform_equal}) and (\ref{poldef}). By writing
Eq.~(\ref{new_flow_law}) in short as
$\mathbf{D}=\mathbf{S}/(2\eta)$ (where $\eta$ is the shear
viscosity of the flow law), we obtain for the crystalline
deformability $S^{\ast}$
\begin{eqnarray}
  S^{\ast}
  &=& 5\,\frac{(\mathbf{S}\cdot\mathbf{n})^2
               -((\mathbf{S}\cdot\mathbf{n})\cdot\mathbf{n})^2}
               {\mathrm{tr}(\mathbf{S}^2)}
  \nonumber\\[1ex]
  &=&
      5\,\frac{(2\eta\mathbf{D}\cdot\mathbf{n})^2
               -((2\eta\mathbf{D}\cdot\mathbf{n})\cdot\mathbf{n})^2}
               {\mathrm{tr}[(2\eta\mathbf{D})^2]}
  \nonumber\\[1ex]
  &=& 5\,\frac{(\mathbf{D}\cdot\mathbf{n})^2
               -((\mathbf{D}\cdot\mathbf{n})\cdot\mathbf{n})^2}
               {\mathrm{tr}(\mathbf{D}^2)}
  \;=\; D^{\ast}\,,
\end{eqnarray}
so that Eq.~(\ref{deform_equal}) is proven. By applying the
averaging operator (\ref{Gammafunc3}) to this result, we find
immediately
\begin{equation}
  S = D\,,
\end{equation}
which is the assertion of Eq.~(\ref{poldef}). Hence, we can
replace $S$ by $D$ in Eq.~(\ref{inv_new_flow_law_1}), which
yields the inverted anisotropic flow law
\begin{equation}
  \mathbf{S} = A^{-1/n}\,\hat{E}^{-1/n}(D)\,
               \left(\frac{\mathrm{tr}(\mathbf{D}^2)}{2}
               \right)^{-(1-1/n)/2}
               \mathbf{D}\,.
  \label{inv_new_flow_law}
\end{equation}

\section{On the anisotropy of the CAFFE flow law}
\label{sect_aniso_glen}

In Sections~\ref{ssect_aniso_flow_law} and
\ref{ssect_aniso_flow_law_inv} we have given the generalization
of Glen's flow law considered in the CAFFE model. The
anisotropy of the CAFFE flow law has recently been analyzed
mathematically by \citet{faria08} through the derivation of the
symmetry group of the CAFFE model. In this section, we prove in
a more direct way that the CAFFE flow is anisotropic despite
the collinearity between the tensors $\mathbf{S}$ and
$\mathbf{D}$, and give examples in order to justify this
choice.

\subsection{CAFFE anisotropy in the context of material theory}

In the context of constitutive theory, the definition of
isotropy states that any rotation of the body in question does
not alter its material response. Mathematically speaking, this
means invariance of the material functions (or functionals) to
arbitrary orthogonal transformations $\mathbf{P}$ of an
undistorted configuration $\kappa$ \citep[e.g.][p.~86]{liu02},
so that the symmetry group $\mathcal{G}$ of the material
contains the entire group of orthogonal transformations O(3)
[$\mathcal{G}\supseteq{}\mathrm{O(3)}$]. Anisotropy is the
logical opposite: for at least one orthogonal transformation
the invariance does not hold, so that the symmetry group does
not contain the entire group of orthogonal transformations
[$\mathcal{G}\not\supseteq{}\mathrm{O(3)}$].

\begin{figure}[htb]
  \centering
  \includegraphics[scale=1.0]{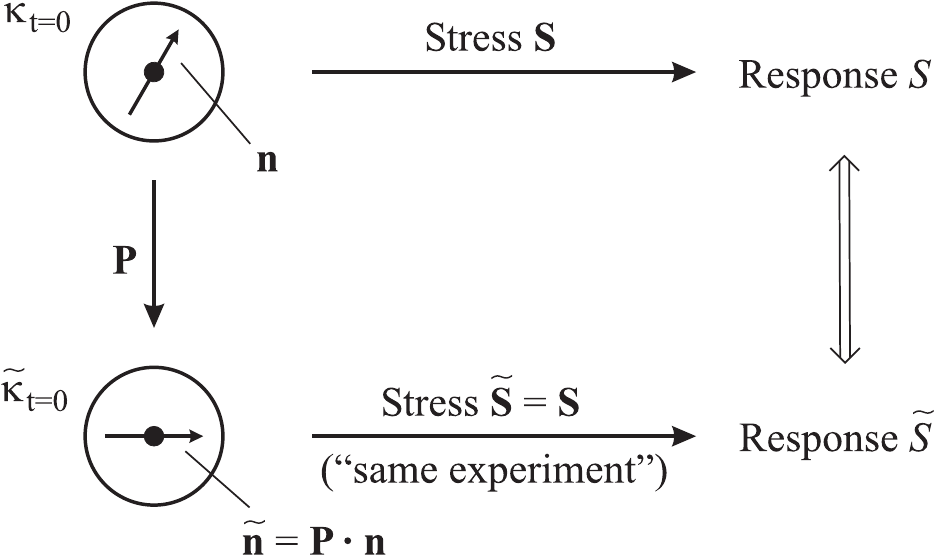}
  \caption{Anisotropy of the CAFFE flow law: If the same stress
  ($\tilde{\mathbf{S}}=\mathbf{S}$) is applied to two rotated
  initial configurations ($\kappa_{t=0}$, $\tilde{\kappa}_{t=0}$),
  the responses $S$ and $\tilde{S}$ are different in general.}
  \label{fig_anisotropy}
\end{figure}

By construction, the anisotropy of the CAFFE flow law
(\ref{new_flow_law}) must be contained in the enhancement
factor $\hat{E}(S)$ via the polycrystal deformability $S$. So
let us assume that, at the initial time $t=0$, the initial
configuration $\kappa_{t=0}$ is given by an unloaded ice
specimen with the OMD
$\varrho^{\ast}(\mathbf{x},\mathbf{n},0)$. At $t=0^+$, it is
subjected to the stress $\mathbf{S}$, and, according to
Eqs.~(\ref{stress_deformability}) and (\ref{Dcdefinition2}),
the resulting polycrystal deformability is
\begin{equation}
  S = \frac{5}{\varrho\,\mathrm{tr}(\mathbf{S}^2)}
      \int_{\mathcal{S}^2} \varrho^{\ast} (\mathbf{n})
                 \left[(\mathbf{S}\cdot\mathbf{n})^2
                  -((\mathbf{S}\cdot\mathbf{n})\cdot\mathbf{n})^2\right]\,
                 \mathrm{d}^2 n\,,
  \label{stress_deformability_unrot}
\end{equation}
where, for simplicity of notation and in the rest of this
session, we will omit the dependence of OMD on position and
time. Now let us consider a second initial configuration
$\tilde{\kappa}_{t=0}$ rotated by a proper orthogonal tensor
$\mathbf{P}$ with respect to $\kappa_{t=0}$. The rotated
orientations are given by
\begin{equation}
  \tilde{\mathbf{n}} = \mathbf{P}\cdot\mathbf{n}
  \label{orientation_trafo}
\end{equation}
(Fig. \ref{fig_anisotropy}). The OMD follows the rotation, so
that
\begin{equation}
  \tilde{\varrho}^{\ast}(\tilde{\mathbf{n}}) = \varrho^{\ast}(\mathbf{n})
  \quad  \stackrel{(\ref{orientation_trafo})}{\Rightarrow}\quad
  \tilde{\varrho}^{\ast}(\tilde{\mathbf{n}}) =
  \varrho^{\ast}(\mathbf{P}^\mathrm{T}\cdot\tilde{\mathbf{n}})\,.
  \label{omd_trafo}
\end{equation}
At $t=0^+$, the rotated configuration is subjected to the
stress $\tilde{\mathbf{S}}$, which is supposed to be the same
as before,
\begin{equation}
  \tilde{\mathbf{S}} = \mathbf{S}
  \label{stress_trafo}
\end{equation}
(Fig. \ref{fig_anisotropy}). The polycrystal deformability with
respect to the rotated configuration is then
\begin{eqnarray}
  \tilde{S}
  &\stackrel{(\ref{stress_deformability_unrot})}{=}&
      \frac{5}{\varrho\,\mathrm{tr}(\tilde{\mathbf{S}}^2)}
      \int_{\mathcal{S}^2} \tilde{\varrho}^{\ast}(\tilde{\mathbf{n}})
                 \left[(\tilde{\mathbf{S}}\cdot\tilde{\mathbf{n}})^2
                  -((\tilde{\mathbf{S}}\cdot\tilde{\mathbf{n}})\cdot\tilde{\mathbf{n}})^2\right]\,
                 \mathrm{d}^2 \tilde{n}
  \nonumber\\[2ex]
  &\stackrel{(\ref{omd_trafo}),\,(\ref{stress_trafo})}{=}&
      \frac{5}{\varrho\,\mathrm{tr}(\mathbf{S}^2)}
      \int_{\mathcal{S}^2} \varrho^{\ast}(\mathbf{P}^\mathrm{T}\cdot\tilde{\mathbf{n}})
                 \left[(\mathbf{S}\cdot\tilde{\mathbf{n}})^2
                  -((\mathbf{S}\cdot\tilde{\mathbf{n}})\cdot\tilde{\mathbf{n}})^2\right]\,
                 \mathrm{d}^2 \tilde{n}\,.
  \label{stress_deformability_rot}
\end{eqnarray}
Let us change the name of the integration variable in the last
integral of Eq.~(\ref{stress_deformability_rot}) from
$\tilde{\mathbf{n}}$ to $\mathbf{n}$,
\begin{equation}\label{stress_deformability_rot2}
  \tilde{S}=\frac{5}{\varrho\,\mathrm{tr}(\mathbf{S}^2)}
      \int_{\mathcal{S}^2} \varrho^{\ast}(\mathbf{P}^\mathrm{T}\cdot\mathbf{n})
                 \left[(\mathbf{S}\cdot\mathbf{n})^2
                  -((\mathbf{S}\cdot\mathbf{n})
                  \cdot\mathbf{n})^2\right]\,
                 \mathrm{d}^2 n\,.
\end{equation}
This is the same as the polycrystal deformability with respect
to $\kappa_{t=0}$ [Eq.~(\ref{stress_deformability_unrot})] for
arbitrary transformations $\mathbf{P}\in\mathrm{O(3)}$ \emph{if
and only if}
$\varrho^{\ast}(\mathbf{n})=\mathrm{const}=\varrho/(4\pi)$. In
this case, the flow law (\ref{new_flow_law}) is isotropic. For
the general case of a non-constant OMD, the deformabilities
(\ref{stress_deformability_unrot}) and
(\ref{stress_deformability_rot2}) are not equal for arbitrary
transformations $\mathbf{P}$, so that the flow law
(\ref{new_flow_law}) is anisotropic, QED. From a mathematical
point of view, Eqs.~(\ref{stress_deformability_unrot}) and
(\ref{stress_deformability_rot2}) make clear that the symmetry
group $\mathcal{G}$ of the material defined by the anisotropic
CAFFE flow law includes the invariance group of orthogonal
transformations that keep the orientation mass density
$\varrho^\ast$ unchanged \citep[e.g.][]{faria08}.

\subsection{Anisotropic behavior for simple shear}

Let us illustrate the anisotropic behavior of the CAFFE flow
law by a simple example. For the vertical single maximum fabric
\begin{equation}
  \varrho^\ast(\mathbf{x},\mathbf{n},t)
  = \varrho(\mathbf{x},t)\,\delta(\mathbf{n}-\mathbf{e}_3)
\end{equation}
and the simple shear deformation
\begin{equation}
  \mathbf{L}
  = \left(
    \begin{array}{ccc}
      0 & 0 & \gamma \\
      0 & 0 & 0 \\
      0 & 0 & 0
    \end{array}
  \right)
  \quad\Rightarrow\quad
  \mathbf{D}
  = \left(
    \begin{array}{ccc}
      0 & 0 & \frac{\gamma}{2} \\
      0 & 0 & 0 \\
      \frac{\gamma}{2} & 0 & 0
    \end{array}
  \right)\,,
  \label{simple}
\end{equation}
we find
\begin{equation}
  \frac{\varrho^\ast}{\varrho}=\delta ( \mathbf{n}-\mathbf{e}_3 )\,,
  \quad \mathbf{S}=\left(
  \begin{array}{ccc}
    0 & 0 & \tau \\
    0 & 0 & 0 \\
    \tau & 0 & 0
  \end{array}
  \right)\,,
  \quad
  S^\ast(\mathbf{e}_3)=\frac{5}{2}
  \quad\Rightarrow\quad
  S=\frac{5}{2}\,.
\label{simple_vsm}
\end{equation}
Thus, the $x$-$z$ component of the flow law
(\ref{new_flow_law}) yields
\begin{equation}
  \label{Eq_SS}
  \gamma = 2 A \hat{E}(\mbox{$\frac{5}{2}$}) \, \tau^n
         = 2 A E_\mathrm{max} \tau^n\,,
\end{equation}
where the explicit definition of the function $\hat{E}(S)$
[Eq.~(\ref{an_glen5})] has been used.

Now we rotate the sample around the direction $\mathbf{e}_2$ by
45$^\circ$ and keep the experimental apparatus fixed [apply the
same stress in the sense of Eq.~(\ref{stress_trafo})]. In this
case, the OMD changes as follows,
\begin{equation}
\label{single_rotated}
  \varrho^\ast=\varrho \,\, \delta (
  \mathbf{n}-\mathbf{e}_{13}),
  \qquad
  \mathbf{e}_{13} = \left(
  \begin{array}{c}
    \frac{1}{\sqrt{2}}  \\
    0 \\
    \frac{1}{\sqrt{2}}
  \end{array}
  \right)\,,
\end{equation}
while the state of stress (\ref{simple_vsm})$_2$ is unchanged.
The species deformability is equal to zero for the orientation
$\mathbf{e}_{13}$,
\begin{equation}
  S^\ast(\mathbf{e}_{13}) = 0
  \quad\Rightarrow\quad
  S=0\,.
\end{equation}
It follows that the stretching tensor evaluated with the flow
law (\ref{new_flow_law}) yields the shear rate
\begin{equation}
  \label{Eq_SS2}
  \gamma
  = 2 A \hat{E}(0) \, \tau^n
  = 2 A E_\mathrm{min} \tau^n\,.
\end{equation}
Since $E_\mathrm{min}\ll{}E_\mathrm{max}$, the shear rate of
Eq.~(\ref{Eq_SS2}) is much smaller than that of
Eq.~(\ref{Eq_SS}). In other words, the material response of the
ice specimen has changed considerably due to the $45^\circ$
rotation of its initial configuration. This fulfills clearly
the criterion for an anisotropic material.

\subsection{Isotropic, anisotropic, collinear and non-collinear flow laws}

On the one hand, the classical Glen flow law
\begin{equation}
  \mathbf{D}
  = A\,\left(\frac{\mathrm{tr}(\mathbf{S}^2)}{2}\right)^{(n-1)/2}
     \,\mathbf{S}\,,
  \label{glen_iso}
\end{equation}
which results from the CAFFE flow law (\ref{new_flow_law}) by
setting $\hat{E}(S)\equiv{}1$, is isotropic and collinear with
respect to the tensors $\mathbf{D}$ and $\mathbf{S}$. On the
other hand, many anisotropic flow laws published so far relate
$\mathbf{D}$ and $\mathbf{S}$ by tensor quantities
\citep{lliboutry93, azuma_94, mangeney_etal_96,
svendsenhutter96, goderthutter98, thorsteinsson_01,
Morland2003, gillet_chaulet_etal_05}, thus giving up the
collinearity between $\mathbf{D}$ and $\mathbf{S}$.

This often leads to the misconception (at least in the
glaciological community) that isotropic flow laws must be
collinear and anisotropic flow laws must be non-collinear with
respect to $\mathbf{D}$ and $\mathbf{S}$. However, this is not
the case. As we have seen above, the CAFFE flow law is
anisotropic, but collinear. Conversely, a non-collinear flow
law is not necessarily anisotropic. An example is the general
Reiner-Rivlin flow law for isotropic viscous fluids
\citep[e.g.][p.~109]{liu02}. For the incompressible case, it
reads
\begin{equation}
  \mathbf{S}
  = \alpha_1 \mathbf{D}
    + \alpha_2
      \left(
        \mathbf{D}^2
        - \frac{\mathrm{tr}\,(\mathbf{D}^2)}{3}\,\mathbf{I}
      \right)\,,
  \label{non_coll}
\end{equation}
where $\alpha_1$ and $\alpha_2$ are material parameters.
Provided that $\alpha_2\ne{}0$, this flow law is evidently
non-collinear. So we highlight that
iso\-tro\-py/an\-iso\-tro\-py and
col\-li\-nea\-rity/non-col\-li\-nea\-rity are two entirely
different concepts, and all four possible combinations can be
realized.

The disadvantage of using a collinear anisotropic flow law is
that, for a given fabric and a given state of stretching, we
select a single viscosity of the polycrystal that is the same
for every components of the stress deviator. However, in
reality, for complex states of stretching (superposition of
compression and shear etc.) different directions will show
different degrees of softening or hardening. This shortcoming
is a tribute to the simple formulation with a scalar
enhancement factor, which allows to set up the flow law with
only two well-known parameters ($E_\mathrm{min}$,
$E_\mathrm{max}$).

\section{Conclusion}
\label{sect_conclusion}

We have presented a constitutive model for the dynamics of
large polar ice masses. This CAFFE model consists of an
anisotropic generalization of Glen's flow law based on a scalar
enhancement factor, and a fabric evolution equation based on an
orientation mass balance. The latter arises from the framework
of Mixtures with Continuous Diversity and uses the orientation
mass density as the variable which describes the anisotropic
fabric. Three constitutive quantities have been introduced,
namely the orientation transition rate due to grain rotation,
the orientation flux and the specific mass production rate.
They have been linked to the physical processes of grain
rotation, rotation recrystallization (polygonization) and grain
boundary migration (migration recrystallization), respectively.
The anisotropy of the CAFFE flow law has been proven, and in
that context it has been emphasized that
iso\-tro\-py/an\-iso\-tro\-py and
col\-li\-nea\-rity/non-col\-li\-nea\-rity (between the stress
and stretching tensors) must be clearly separated. Some
applications of the CAFFE model to simple deformation states
have been discussed (in the appendix), for which analytical
solutions could be obtained, and which could easily be checked
for their physical plausibility and consistency with
observations.

Due to its relative simplicity, the CAFFE model is suitable for
implementation in ice-flow models. This has already been done
by \citet{Seddik2008} for a one-dimensional model of the site
of the EPICA ice core at Kohnen Station in Dronning Maud Land,
East Antarctica \citep{epica_06}, and by \citet{seddik_08} and
\citet{Seddik2009} for the three-dimensional, full-Stokes model
Elmer/Ice in order to simulate the ice flow in the vicinity
within 100~km around the Dome Fuji drill site
\citep{motoyama_07} in central East Antarctica.

\begin{small}

\section*{Acknowledgements}

The authors would like to thank Kolumban Hutter and Leslie W.\
Morland for many productive discussions. Comments of the
scientific editor Wolfgang M\"uller and an anonymous reviewer
helped considerably to improve the structure and clarity of the
manuscript. This work was supported by a Grant-in-Aid for
Creative Scientific Research (No.\ 14GS0202) from the Japanese
Ministry of Education, Culture, Sports, Science and Technology,
by a Grant-in-Aid for Scientific Research (No.\ 18340135) from
the Japan Society for the Promotion of Science, and by a grant
(Nr.\ FA 840/1-1) from the Priority Program SPP-1158 of the
Deutsche Forschungsgemeinschaft (DFG).

\begin{appendix}

\section{Examples for the evolution of the orientation mass density}
\label{sect_omd_evol}

\subsection{Evolution due to grain rotation}
\label{ssect_omd_evol_1}

The deck-of-cards deformation mechanism (grain rotation)
implies that the $c$-axis of a crystallite in the
polycrystalline aggregate rotates towards the axes of
compression and away from that of extension. This is
illustrated graphically in Fig.~\ref{rotrec} for the case of
rotated pure shear (simple shear) and described mathematically
by Eq.~(\ref{constpart_u}), provided that the constitutive
parameter $\iota$ is positive. Equation~(\ref{constpart_u})
fulfills the principle of material frame indifference. Thus, if
the rules for compression and for extension are satisfied, then
the rules for simple shear are a direct consequence. Here we
give some simple examples in which this can explicitly be seen.

We use a Cartesian frame of reference for which the orientation
$\mathbf{n}$ of the $c$-axis is parameterized by
Eq.~(\ref{sphericalvectorial}). For uniaxial vertical
compression (transversely isotropic horizontal extension), the
stretching tensor is
\begin{equation}
  \mathbf{D}=\left(
  \begin{array}{ccc}
  \frac{\varepsilon}{2} & 0 & 0 \\
  0 & \frac{\varepsilon}{2} & 0 \\
  0 & 0 & -\varepsilon
  \end{array}
  \right)
  = \frac{1}{2} \, \varepsilon \,
    \mathbf{e}_1\,\mathbf{e}_1
    + \frac{1}{2}\, \varepsilon \, \mathbf{e}_2 \, \mathbf{e}_2
    - \varepsilon \, \mathbf{e}_3\,\mathbf{e}_3\,,
  \label{purevertical}
\end{equation}
where $\varepsilon>0$ holds. From
Eqs.~(\ref{sphericalvectorial}), (\ref{constpart_u}) and
(\ref{purevertical}) we derive the explicit form of the
orientation transition rate due to grain rotation,
\begin{eqnarray}
  \mathbf{u}_\mathrm{gr}^{\ast}
  &=& -\frac{3}{4} \, \iota \, \varepsilon \sin 2\theta
  \left(
  \begin{array}{c}
  \cos \varphi \cos \theta  \\
  \sin \varphi \cos \theta  \\
  -\sin \theta
  \end{array}
  \right)
  \nonumber\\[1ex]
  &=& - \frac{3}{4} \, \iota \, \varepsilon \sin 2\theta \,
  (\cos\varphi\cos\theta\,\mathbf{e}_1 + \sin\varphi\cos\theta\,\mathbf{e}_2
  - \sin\theta\,\mathbf{e}_3)\,.
  \label{u3}
\end{eqnarray}
The direction of $\mathbf{u}_\mathrm{gr}^{\ast}$ is coherent
with the rules of Fig.~\ref{rotrec}a (see also
Fig.~\ref{ssquareandu}b): The third component
$\mathbf{u}_\mathrm{gr}^{\ast}\cdot\mathbf{e}_3$ is positive
when $\theta<\pi/2$ and negative when $\theta>\pi/2$. If the
crystallites are in the plane spanned by $\mathbf{e}_1$ and
$\mathbf{e}_3$ ($\varphi=0$), Eq.~(\ref{u3}) simplifies to
\begin{equation}
  \varphi = 0
  \quad\Rightarrow\quad
  \mathbf{u}_\mathrm{gr}^{\ast} = \frac{3}{4} \, \iota \, \varepsilon \sin 2\theta
  \left(
  \begin{array}{c}
  -\cos\theta \\
  0 \\
  \sin\theta
  \end{array}
  \right)\,,
  \label{u4}
\end{equation}
and if they are in the plane spanned by $\mathbf{e}_2$ and
$\mathbf{e}_3$ ($\varphi=\pi/2$), we find
\begin{equation}
  \varphi = \frac{\pi}{2}
  \quad\Rightarrow\quad
  \mathbf{u}_\mathrm{gr}^{\ast} = \frac{3}{4} \, \iota \, \varepsilon \sin 2\theta
  \left(
  \begin{array}{c}
  0 \\
  -\cos \theta  \\
  \sin \theta
  \end{array}
  \right)\,.
  \label{u5}
\end{equation}
It is worth noting that the norm of
$\mathbf{u}_\mathrm{gr}^{\ast}$ which results from
Eqs.~(\ref{u3}), (\ref{u4}) or (\ref{u5}) shows the explicit
dependence on the Schmidt factor $\sin{}2\theta$ that
\citet{azumahigashi85} recognized experimentally. The presence
of the Schmidt factor guarantees that crystallites with
vertical and horizontal orientations do not rotate, while those
at orientations $45^{\circ}$ off the vertical show maximum
rotation.

If the uniaxial compression is along the first ($\mathbf{e}_1$)
instead of the third $(\mathbf{e}_3)$ axis of the frame of
reference, the orientation transition rate due to grain
rotation takes the form
\begin{equation}
  \mathbf{D} = \left(
  \begin{array}{ccc}
  -\varepsilon & 0 & 0 \\
  0 & \frac{\varepsilon}{2} & 0 \\
  0 & 0 & \frac{\varepsilon}{2}
  \end{array}
  \right)
  \quad\Rightarrow\quad
  \mathbf{u}_\mathrm{gr}^{\ast} = \iota \, \varepsilon
  \left(
  \begin{array}{c}
  \sin \theta \cos \varphi
  +B\left(\theta ,\varphi\right)\sin \theta \cos \varphi \\
  -\frac{1}{2}\sin \theta \sin \varphi +B\left(\theta
  ,\varphi\right)\sin \theta \sin \varphi
  \\
  -\frac{1}{2}\cos \theta +B\left(\theta ,\varphi\right) \cos \theta
  \end{array}
  \right)\,,
  \label{purehorizontal1}
\end{equation}
where
\begin{equation}
  B\left(\theta ,\varphi\right)\equiv - \sin \theta \cos \varphi
  \sin \theta \cos \varphi + \frac{1}{2}\sin \theta \sin \varphi
  \sin \theta \sin \varphi + \frac{1}{2}\cos \theta \cos \theta\,.
\end{equation}
The difference to Eq.~(\ref{u3}) arises only because the
spherical coordinate system, which underlies the representation
of the unit vector $\mathbf{n}$ in
Eq.~(\ref{sphericalvectorial}), is more convenient for the
vertical compression (\ref{purevertical}) than for the
horizontal compression (\ref{purehorizontal1})$_1$.

For a planar elongation (or pure shear) state of deformation,
\begin{equation}
  \mathbf{D}=\left(
  \begin{array}{ccc}
  \varepsilon & 0 & 0 \\
  0 & 0 & 0 \\
  0 & 0 & -\varepsilon
  \end{array}
  \right)
  = \varepsilon\,\,\mathbf{e}_1\,\mathbf{e}_1
    - \varepsilon\,\,\mathbf{e}_3\,\mathbf{e}_3 \,,
  \label{pureverticalstrong}
\end{equation}
the orientation transition rate due to grain rotation results
from Eqs.~(\ref{sphericalvectorial}), (\ref{constpart_u}) and
(\ref{pureverticalstrong}) as
\begin{equation}
  \mathbf{u}_\mathrm{gr}^{\ast}=\iota \, \varepsilon\left(
  \begin{array}{c}
  ( - 1 + \sin \theta \cos \varphi \sin \theta \cos \varphi -
  \cos \theta
  \cos \theta ) \, \sin \theta \cos \varphi  \\
  0 \\
  \sin \theta \sin \theta \,
  ( 1 + \cos \varphi \cos \varphi ) \, \cos \theta
  \end{array}
  \right)\,,
  \label{u13}
\end{equation}
which, in the plane spanned by $\mathbf{e}_1$ and
$\mathbf{e}_3$, is
\begin{equation}
  \varphi
  = 0\quad \Rightarrow \quad \mathbf{u}_\mathrm{gr}^{\ast}=\iota \,
  \varepsilon \sin 2\theta\left(
  \begin{array}{c}
  - \cos \theta  \\
  0 \\
  \sin \theta
  \end{array}
  \right)\,.
  \label{u14}
\end{equation}
This is larger by the factor $4/3$ compared to the orientation
transition rate of Eq.~(\ref{u4})$_2$. The reason for this
difference is that the component $D_{22}$ does not contribute
to grain rotation in the plane spanned by $\mathbf{e}_1$ and
$\mathbf{e}_3$, which makes the orientation transition rate in
the case of Eq.~(\ref{u14}) (where $D_{22}=0$) faster than in
the case of Eq.~(\ref{u4}) (where $D_{22}=\varepsilon/2$).

For the simple shear situation of Eq.~(\ref{simple}), which is
illustrated in Fig.~\ref{rotrec}b, we find for the orientation
transition rate due to grain rotation
\begin{equation}
  \mathbf{u}_\mathrm{gr}^{\ast}
  = \iota \, \frac{\gamma}{2}
  \left(
  \begin{array}{c}
  \cos \theta \, ( 2 \sin^2 \theta \cos^2 \varphi -1 )  \\
  \frac{1}{2} \sin \theta \sin 2 \theta \sin 2 \varphi \\
  \sin \theta \cos \varphi \cos 2 \theta
  \end{array}
  \right)\,,
\end{equation}
which, in the plane spanned by $\mathbf{e}_1$ and
$\mathbf{e}_3$, gives
\begin{equation}
  \varphi = 0
  \quad\Rightarrow\quad
  \mathbf{u}_\mathrm{gr}^{\ast}
  = \iota \, \frac{\gamma}{2} \cos 2\theta
  \left(
  \begin{array}{c}
  -\cos \theta  \\
  0 \\
  \sin \theta
  \end{array}
  \right)\,.
  \label{u14bis}
\end{equation}
The direction of $\mathbf{u}_\mathrm{gr}^{\ast}$ is once more
consistent with the rules of Fig.~\ref{rotrec}b (see also
Fig.~\ref{ssquareandu}b). For instance, for crystallites
oriented upward within $\theta<\pi/4$ and a positive shear rate
$\gamma>0$, the component of $\mathbf{u}_\mathrm{gr}^{\ast}$
along $\mathbf{e}_1$ is negative.

\subsection{Evolution due to recrystallization}
\label{ssect_omd_evol_2}

We now give some examples for the recrystallization term of
Eq.~(\ref{Gammafunc2}) for standard deformation situations, and
also provide a model of the experiments by
\citet{budd_jacka_89} for uniaxial compression with isotropic
horizontal extension.

For the case of pure shear rate as defined in
Eq.~(\ref{pureverticalstrong}), crystallites oriented
vertically have a vanishing species deformability,
\begin{equation}
  \mathbf{n} = \mathbf{e}_{3}
  \quad \Rightarrow \quad
  D^\ast = 5\,\frac{(\mathbf{D}\cdot\mathbf{n})^2
  -\left((\mathbf{D}\cdot\mathbf{n})
  \cdot \mathbf{n}\right)^2}{\mathrm{tr}(\mathbf{D}^2)}
  = 5\,\frac{\varepsilon^2-\varepsilon^2}{\frac{3}{2}\varepsilon^2}
  = 0\,,
  \label{zerostretch}
\end{equation}
while crystallites inclined by $45^\circ$ off the vertical
towards $\mathbf{e}_1$ have the maximum deformability,
\begin{equation}
  \mathbf{n}
  = \mathbf{e}_{13}
  = \left(
  \begin{array}{c}
  \frac{1}{\sqrt{2}}  \\
  0 \\
  \frac{1}{\sqrt{2}}
  \end{array}
  \right)
  \quad \Rightarrow \quad
  D^\ast = 5\,\frac{(\mathbf{D}\cdot\mathbf{n})^2
  -\left((\mathbf{D}\cdot\mathbf{n}) \cdot
  \mathbf{n}\right)^2}{\mathrm{tr}(\mathbf{D}^2)}
  = 5\,\frac{\varepsilon^2-0}{2 \varepsilon^2} =\frac{5}{2}\,.
  \label{halfstretch}
\end{equation}
For general anisotropic fabrics, the averaged deformability
$\langle{}D^{\ast}\rangle$ is between these extremes. It
follows from Eq.~(\ref{Gammafunc2}) that the favourably
oriented crystals with $\mathbf{n}=\mathbf{e}_{13}$ will grow
($\Gamma^\ast>0$) and the unfavourably oriented ones with
$\mathbf{n}=\mathbf{e}_{3}$ will shrink ($\Gamma^\ast<0$). This
is the physically expected behaviour.

The experiments by \citet{budd_jacka_89} were carried out under
the deformation regime of uniaxial compression with isotropic
horizontal extension, as specified by Eq.~(\ref{purevertical}).
By using the parameterization (\ref{sphericalvectorial}) for
general orientations $\mathbf{n}$, we compute the species
deformability (\ref{Dcdefinition}) as
\begin{equation}
  \label{shmidtstretch}
  D^\ast = \frac{15}{2} \sin^2 \theta \cos^2 \theta\,.
\end{equation}
If at the initial time $t=0$, the OMD is random
($\varrho^\ast=\varrho/4\pi$, isotropic fabric), then
\begin{equation}
  D = \langle D^{\ast} \rangle
    = \int_ {\mathcal{S}^{2}} \frac{\varrho^{\ast}}{\varrho} \,
      D^{\ast} \, \mathrm{d}^2 n
    = \int_{0}^{2 \pi} \frac{1}{4 \pi}\,\mathrm{d}\varphi
      \int_{0}^{\pi} \frac{15}{2} \,
      \sin^2\theta \cos^2\theta\,
      \sin\theta\,\mathrm{d}\theta = 1\,,
  \label{dstar_av}
\end{equation}
where
$\mathrm{d}^2{}n=\sin\theta\,\mathrm{d}\theta\,\mathrm{d}\varphi$
and usual integration rules have been used. The specific mass
production rate which results from Eqs.~(\ref{Gammafunc2}) and
(\ref{dstar_av}) is
\begin{equation}
  \Gamma^\ast(\theta)
  = \hat{\Gamma} \left( \frac{15}{2} \sin^2\theta \cos^2\theta - 1 \right)
  = \frac{15}{8} \, \hat{\Gamma} \, \sin^2 2\theta - \hat{\Gamma}\,.
  \label{gammasimple}
\end{equation}
Consequently, crystallites with
$\theta\in\left(\frac{1}{2}\arcsin\sqrt{\frac{8}{15}},\,
\frac{\pi}{2}-\frac{1}{2}\arcsin\sqrt{\frac{8}{15}}\right)$
grow [$\Gamma^\ast(\theta)>0$], while the others shrink
[$\Gamma^\ast(\theta)<0$], and an anisotropic fabric evolves.
If we do not consider grain rotation and rotation
recrystallization, then, asymptotically for $t\to\infty$, we
will obtain
\begin{equation}
  \varrho^\ast
  = \frac{\varrho}{2\pi\sin\theta_0} \,
    \delta ( \theta - \theta_0 ),
    \quad \mbox{with} \;\, \theta_0 = 45^\circ \,,
  \label{dirac}
\end{equation}
where $\delta$ is the Dirac delta function.
Equation~(\ref{dirac}) is the mathematical representation of a
girdle fabric \citep[see e.g.][]{placidihutter06_zamp} in which
all the crystallites are inclined by $45^\circ$ with respect to
the vertical. If grain rotation is superimposed, these
crystallites will experience an additional rotation towards the
compression axis $\mathbf{e}_3$ [in accordance with
Eq.~(\ref{u14bis}) or Fig.~\ref{rotrec}a], so that the small
girdle fabric observed by \citet{jackabudd89} is deduced.

Another interesting example is the rotated pure shear (simple
shear) regime of Eq.~(\ref{simple}) for general orientations
$\mathbf{n}$ represented by Eq.~(\ref{sphericalvectorial}). The
crystal deformability is computed for this case as
\begin{equation}
  D^{\ast}
  = \frac{5}{2} \left[ \cos^2\varphi \sin^2\theta
    + \cos^2\theta \,
    ( 1 - 4 \cos^2\varphi \sin^2\theta ) \right]\,.
\label{dstar_simple}
\end{equation}
If at the initial time $t=0$, the OMD is random
($\varrho^\ast=\varrho/4\pi$, isotropic fabric), then, analogue
to Eqs.~(\ref{dstar_av}) and (\ref{gammasimple}), we find
\begin{equation}
  D = 1
  \label{dstar_simple_av}
\end{equation}
and
\begin{equation}
  \Gamma^\ast(\theta,\varphi)
  = \frac{5}{2}\,\hat{\Gamma} \left[ \cos^2\varphi \sin^2\theta
    + \cos^2\theta \,
    ( 1 - 4 \cos^2\varphi \sin^2\theta ) \right]
    - \hat{\Gamma}\,.
  \label{gamma_simple}
\end{equation}
For times $t>0$, an anisotropic fabric evolves, because
crystallites oriented near $\mathbf{n}=\mathbf{e}_1$
($\theta\approx\pi/2$ and $\varphi\approx{}0$) or
$\mathbf{n}=\mathbf{e}_3$ ($\theta\approx{}0$) grow and the
others shrink. Hence, without local rigid body rotation, grain
rotation and rotation recrystallization, asymptotically for
$t\to\infty$ we will obtain the two-maxima fabric
\begin{equation}
  \varrho^\ast
  = \frac{1}{2} \delta\left( \mathbf{n}-\mathbf{e}_1 \right)
    + \frac{1}{2} \delta\left( \mathbf{n}-\mathbf{e}_3 \right)\,.
  \label{twomaxima}
\end{equation}
If grain rotation and rigid body rotation are superimposed, the
fabric reported by \citet{kamb72} results.

\end{appendix}

% \section*{References}

% \bibliographystyle{ralf}
% \bibliography{journals2_ralf,0903.0688v4}

\end{small}

\end{document}